\documentclass[epj,final]{svjour}[2 jan 2009]
\usepackage{cite}
\usepackage{amsmath,amssymb}
\usepackage{graphicx}
\usepackage{times}

\DeclareMathOperator*{\grad}{grad}

\begin{document}
\title{Evaporation induced flow inside circular wells}
\author{Yu.~Yu.~Tarasevich \and I.~V.~Vodolazskaya \and O.~P.~Isakova \and M.~S.~Abdel Latif}
\institute{Astrakhan State University, 20a Tatishchev St.,
Astrakhan, 414056, Russia, \email{tarasevich@aspu.ru}}
\date{}
\abstract{Flow field and height averaged radial velocity inside a
droplet evaporating in an open circular well were calculated for
different modes of liquid evaporation.
 % end of PACS codes
} %end of abstract
\maketitle
\section{Introduction}
\label{sec:intro}
Wide used technique of patterning surfaces with solid particles
utilizes the evaporation of colloidal droplets from a substrate. In
particular, evaporation of liquid samples is a key problem in the
development of microarray technology (including labs-on-a chip),
especially in the case of open reactors~\cite{Rieger}.

\begin{figure}
  \centering
  \includegraphics*[width=0.95\linewidth]{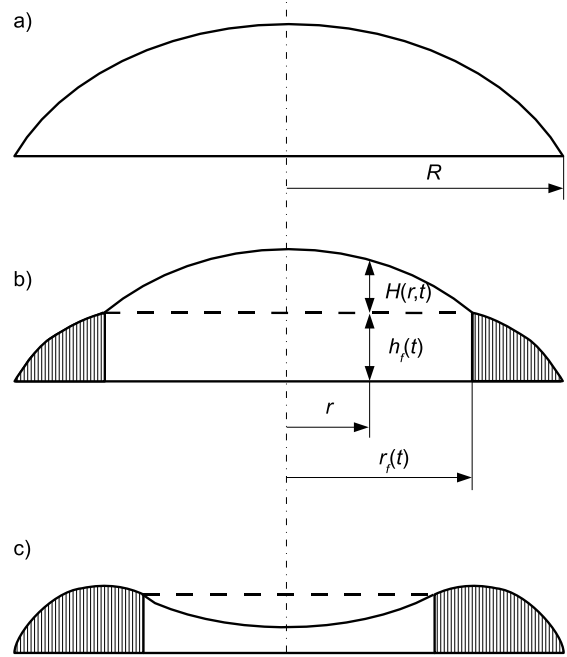}\\
  \caption{Time evolution of the deposit phase growth~\cite{Popov_2005}}\label{fig:popov}
\end{figure}

A plenty of works are devoted to the evaporation of the sessile
droplets~\cite{Deegan_2000,Fischer_2002,Mollaret_2004,Hu_2005}. One
assumes that a droplet has a shape like a spherical cap.
Nevertheless, investigation of the drops with more complex geometry
is of interest. Thus when a high concentrated colloidal droplet
evaporates from a substrate, the solute particles in solution will
form a ring like deposit wall near the edge of the
drop~(Fig.~\ref{fig:popov}). One can consider this case as
evaporation of a liquid drop inside open circular well with vertical
walls formed by gel.

In the paper~\cite{Rieger} the flow field inside a liquid sample in
very thin circular wells was measured. The confidence between
measured velocity~\cite{Rieger} and the estimations obtained with
theory of ring formation~\cite{Deegan_2000} is reasonable.

In this paper, the relationship between hydrodynamical flow inside a
liquid sample evaporating in open circular well and the mode of
evaporation is examined.

We found analytical expressions of height averaged radial velocity
for different modes of evaporation. Velocity field inside the
circular well was found for the particular case of flat air--liquid
interface.

\section{Velocity field inside circular well}\label{sec:velocityfield}
\subsection{Height averaged radial velocity}\label{subsec:hav}

We will consider one particular but interesting case, because of its
practical applications, when a liquid inside a circular well may be
described as a thin film. So, in paper~\cite{Rieger} in circular
wells with a radius of $r_f =100$~$\mu$m and a depth of $h_f =
6$~$\mu$m were investigated. Approximately the same ratio depth to
radius is typical for the drops of biological fluids used for
medical tests~\cite{Savina_1999,Shabalin_2001}.

Moreover, we will suppose that evaporation is a steady state
process. This assumption is valid e.g. for evaporation of the drops
of aqueous solutions under room temperature and normal atmosphere
pressure, i.e. for the typical experimental conditions. Apex
dynamics is rather slow~\cite{Rieger}
$$
H(t) = (7.38 \mathrm{\mu m} - 6.13 \mathrm{\mu m}) - (0.042
\mathrm{\mu m/s})t.
$$
In the cases of medical tests, the typical velocity of drop apex is
approximately 1~mm/h.

Mass conservation gives the following equation for height averaged
radial velocity~\cite{Deegan_2000}
\begin{multline}\label{eq:vr}
 \langle v_r(r,t)\rangle = \\- \frac{1}{\rho r h} \int\limits_0^r  \left( j(r,t) \sqrt{1+\left(
  \frac{\partial  h}{\partial  r}\right)^2}  +  \rho  \frac{\partial
  h}{\partial  t}  \right) r\, dr,
\end{multline}
where $h = h(r,t)$ is the drop shape, $\rho$ is the density of
solution, $j(r,t)$ is the evaporation rate defined as the
evaporative mass loss per unit surface area per unit time.

Considering the thin droplets only ($h(r,t) \ll r_f)$ we will
neglect everywhere $h^2(0,t)$ and $\left(\frac{\partial h}{\partial
r}\right)^2$ in compare with 1. We will utilize approximative
equation for air--liquid interface
\begin{equation}\label{eq:profilethincirc}
    h(r,t) = h(0,t) \left(1 -
    \left(\frac{r}{r_f}\right)^2\right),
\end{equation}
where $r_f$ is radius of circular well.

Since the drop is thin and the contact angle is small, we will use
the expression
\begin{equation}\label{eq:Jthindisc}
    j(r) = \frac{j_0}{\sqrt{1 - \left(\frac{r}{r_f}\right)^2}}
\end{equation}
for the evaporation rate, which has a reciprocal square-root
divergence near the contact line~\cite{Popov_2005}. This expression
can be derived from Laplace equation
(see~\cite{Deegan_2000,Popov_2005} for details). This functional
form for vapor flux is widely used, in particular, it was utilized
in paper~\cite{Rieger}.

Some quantities of interest can be expressed analytically
exploiting~\eqref{eq:Jthindisc}. Velocity of drop apex decrease is
\begin{equation}\label{eq:vthindisc}
   \frac{d h(0,t)}{d t} = - \frac{4 j_0}{\rho}.
\end{equation}
Height averaged radial velocity is
\begin{equation}\label{eq:vrxtdisc}
    \langle v_r(x,t) \rangle = -\frac{j_0  \left( 1 - \sqrt{ 1 - x^2} - 2 x^2 + x^4
    \right)}{\rho x (h_0 + L(0,t) ( 1 - x^2))},
\end{equation}
where  $x = \frac{r}{r_f}$, $L = \frac{h}{r_f}$, $h_0 =
\frac{h_f}{r_f}$. $h_f$ is the height of vertical wall of the well.

Exploiting of Eq.~\ref{eq:Jthindisc} leads to a singularity for both
vapor flux and gradient of the height averaged velocity of liquid
inside a drop at the edge of droplet. A smoothing function may be
used to eliminate physically senseless
divergency~\cite{Cachile_2002}
\begin{equation}\label{eq:Jthindiscsmooth}
    j(x,t) = \frac{j_0}{r_f\sqrt{1 - x^2}}\frac{1 - \exp(- m \sqrt{1 - x^2})}{1 -
    \exp(-m)},
\end{equation}
where $m$ is an adjustable constant. In this case, height averaged
radial velocity can be written analytically
\begin{multline}\label{eq:Jdsanalyt}
\langle v_r(x,t) \rangle = \frac { j_0  }{x( h_0 + L \left( x,t
\right)) m \left( 1-{{\rm e}^{-m}} \right) }\times \\ \left( m\sqrt
{1-x^2}+{{\rm e}^{-m\sqrt {1-{x}^ {2}}}}-m-{{\rm e}^{-m}}-
\phantom{\frac{x^2}{2}} \right.
\\ \left. 2\, \left( 1-m-{{\rm e}^{-m}} \right) \left( 1- \frac{x^2}{2}
\right) x^2 \right),
\end{multline}
where $L (x, t) = \left(  L(0,0) + 4{\frac {{ j_0}\, \left(
1-m-{{\rm e}^{- m}} \right) t}{m \left( 1-{{\rm e}^{-m}} \right) }}
\right) \left( 1- x^2 \right). $

Another evaporative flux functions was proposed
in~\cite{Anderson_1995}
\begin{equation}\label{eq:Jdavis}
j(x,\tau) = \frac{j_0}{K + L(x,\tau)},
\end{equation}
where the constant $K$ measures the degree of non-equilibrium at the
evaporating interface and is related to material properties. $K \to
0$ corresponds to a highly volatile droplet. The limit $K \to
\infty$ corresponds to a nonvolatile droplet. Analytical expression
for height averaged velocity inside circular well can be obtained if
vapor flux is described by Eq.~\ref{eq:Jdavis}
\begin{multline}\label{eq:havanaldavis}
 \langle v_r(x,t) \rangle = \frac{j_0}{2 x \left( h_0 + L(0,t)(1 - x^2) \right) } \times \\
 \times \left( \frac{1}{L(0,t)}
 \ln  \left( 1 - \frac {L (0,t) x^2}{K + h_0 + L(0,t) }\right) -
 \right. \\ \left.
 \frac {x^2 \left( 1 - \frac{x^2}{2} \right) \left( L(0,t) - 2(K + h_0) \right) }{
 \left( K + h_0 \right) ^2} \right),
\end{multline}
where $L(0,t) = 2 (K + h_0) + (L_0 - 2(K + h_0))\exp\left( \frac{j_0
t}{(K + h_0)^2} \right).$

Deegan et al.~\cite{Deegan_2000} demonstrated experimentally, that
if evaporation is greatest at the center of the droplet, a uniform
distribution of colloidal particles remained on the substrate.
Simulations by~\cite{Fischer_2002} confirmed, that as fluid is lost
from the center of the droplet, an inward flow develops to replenish
the evaporated fluid. The evaporative flux function was proposed by
Fischer~\cite{Fischer_2002} to mimic evaporation that is
concentrated at the center of the droplet
\begin{equation}\label{eq:Jgauss}
    j(x,t) = \frac{j_0}{L(0,t)}\mathrm{e}^{-Ax^2},
\end{equation}
where  $A$ is an adjustable constant.

We derived the analytical expression for height averaged velocity
for evaporation function given by Eq.~\ref{eq:Jgauss}
\begin{multline}\label{eq:gauss}
\langle v_r(x,t) \rangle =  \\ \frac{j_0 \left( 1-{{\rm e}^{-Ax^2}}+
\left( 1-{{\rm e}^{-A} } \right) x^2 \left( x^2-2 \right) \right) }{
 2 A x (h_0 + L(x,t)) L(0,t)},
\end{multline}
where
\begin{multline*}
L(x,t) = \\ \left( 1 - x^2 \right)  \left( \sqrt { \left( h_0 + L_0
\right) ^2 - 4 \frac { j_0\, \left( 1 - {\rm e}^ {-A} \right)
t}{A}}- h_0 \right).\end{multline*}

\begin{figure*}
\centering
a) \includegraphics[width=0.4\textwidth]{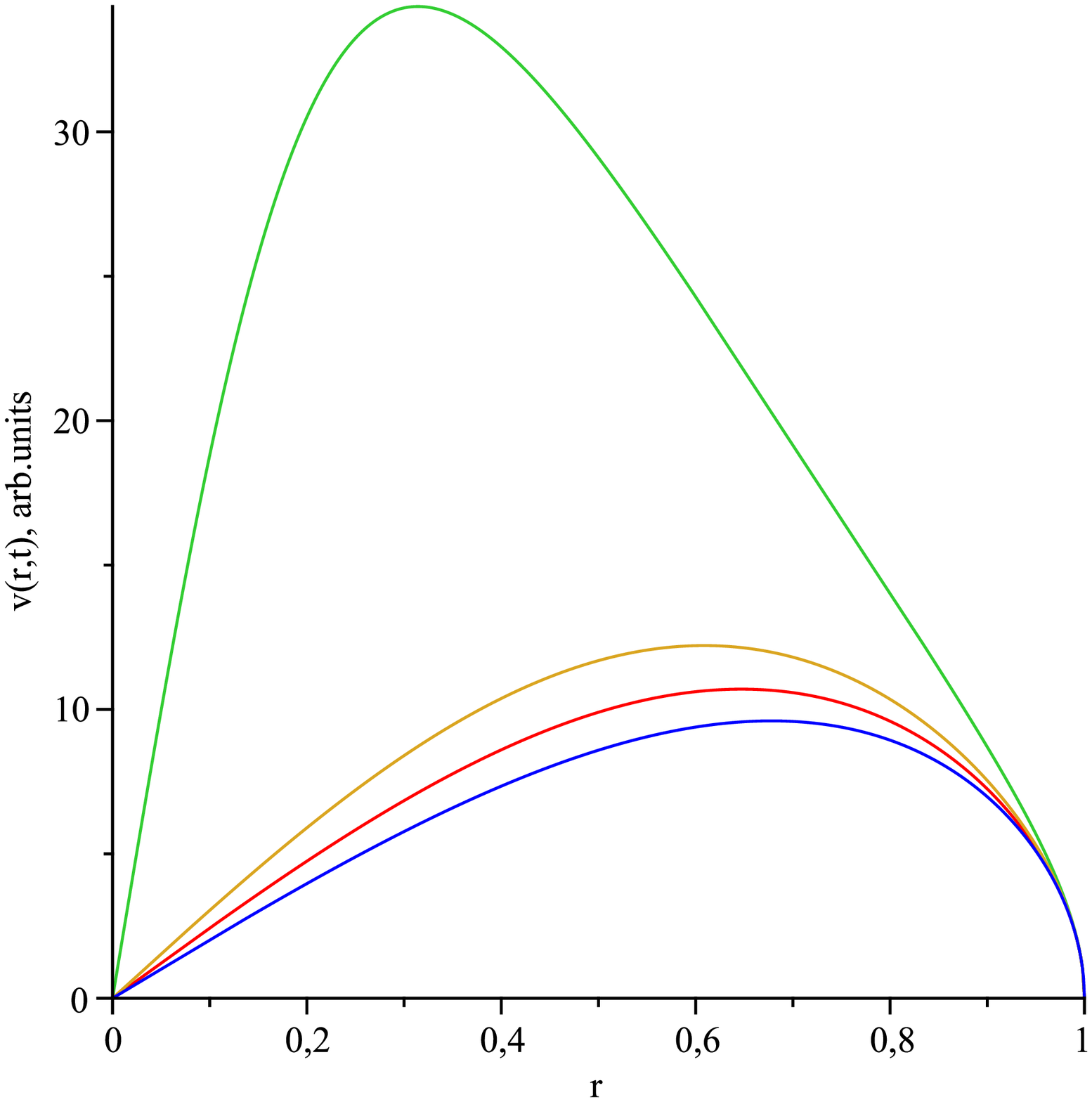}   \hfill  b) \includegraphics[width=0.4\textwidth]{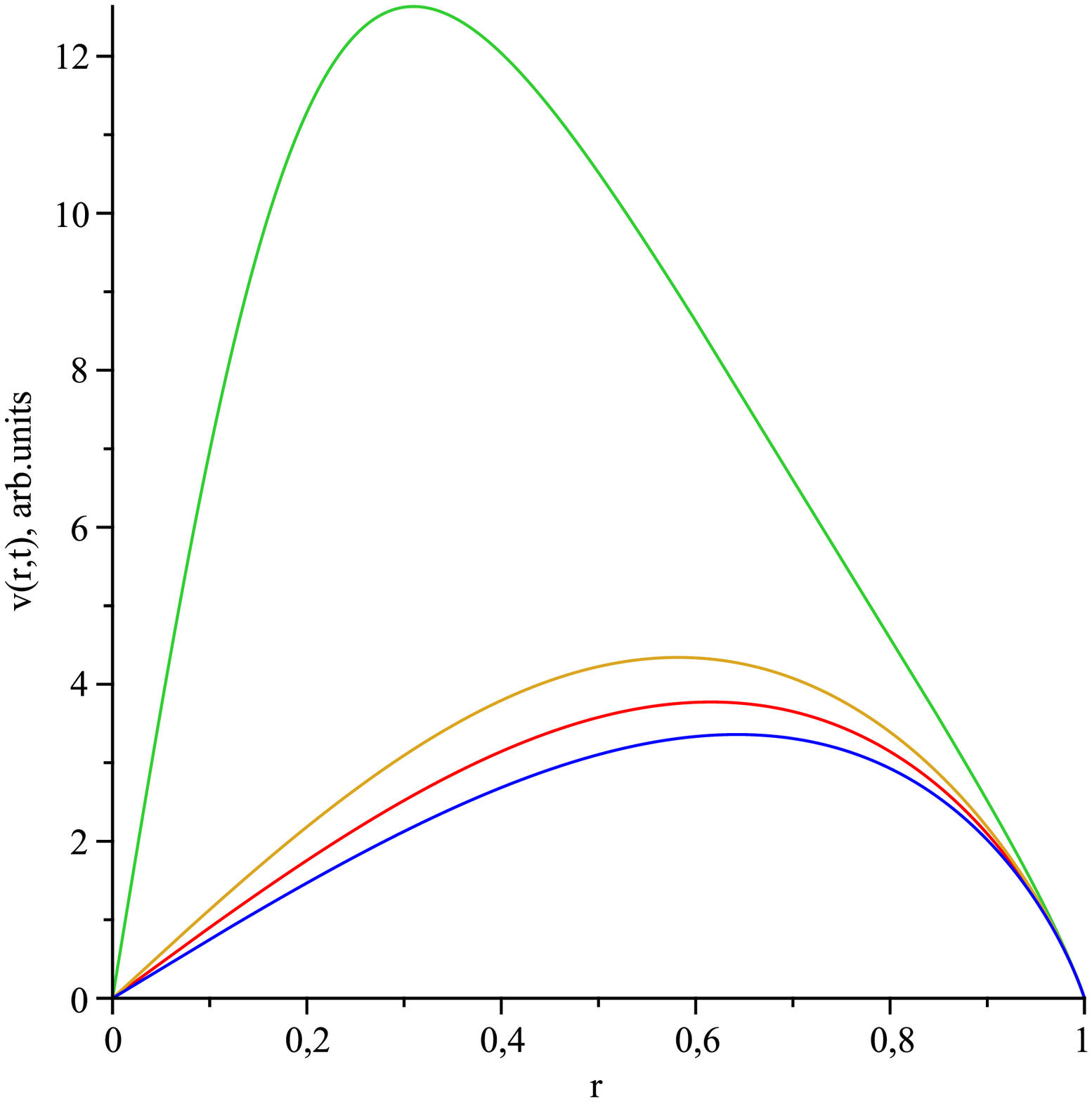}\\
c) \includegraphics[width=0.4\textwidth]{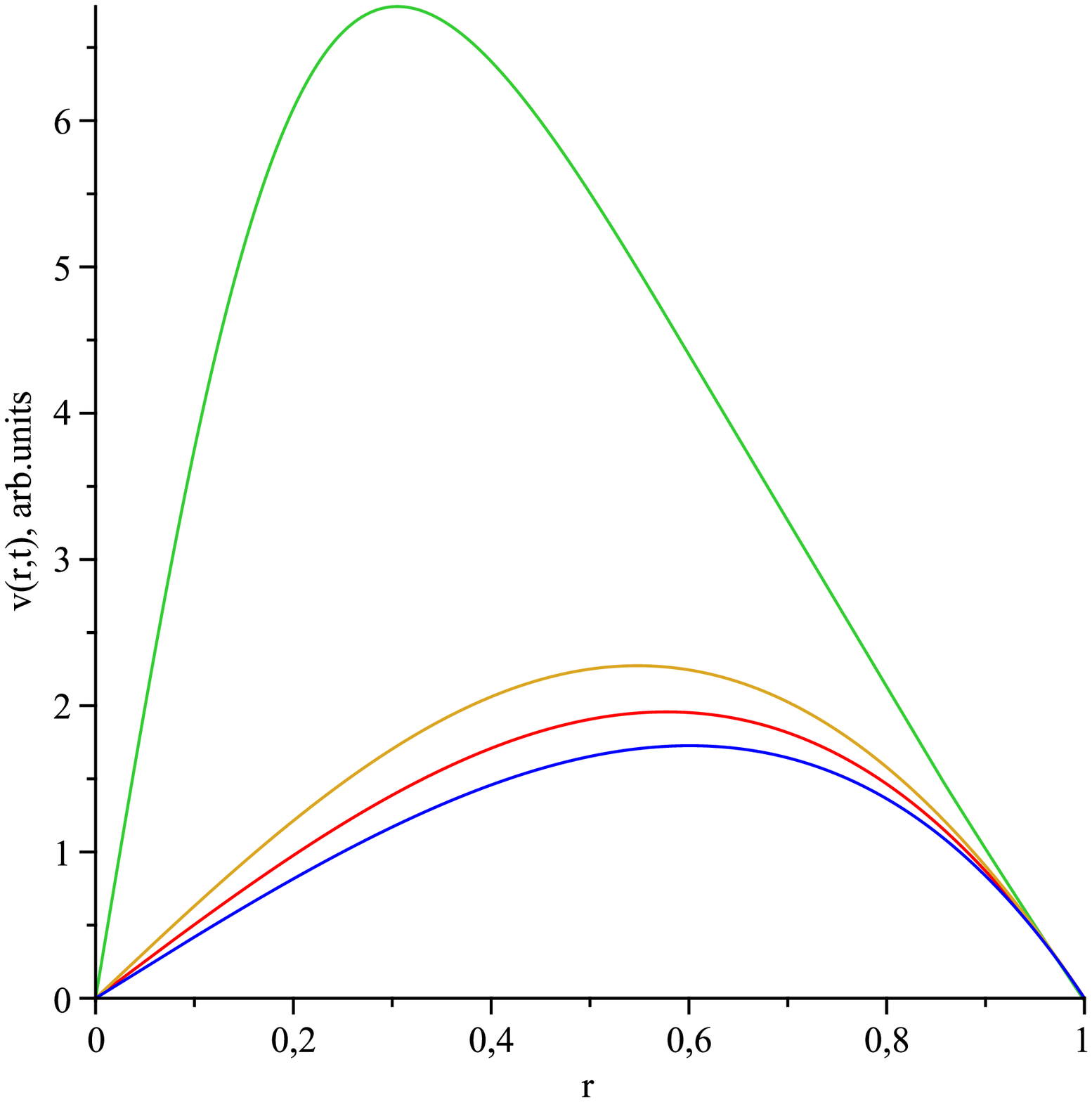}  \hfill  d) \includegraphics[width=0.4\textwidth]{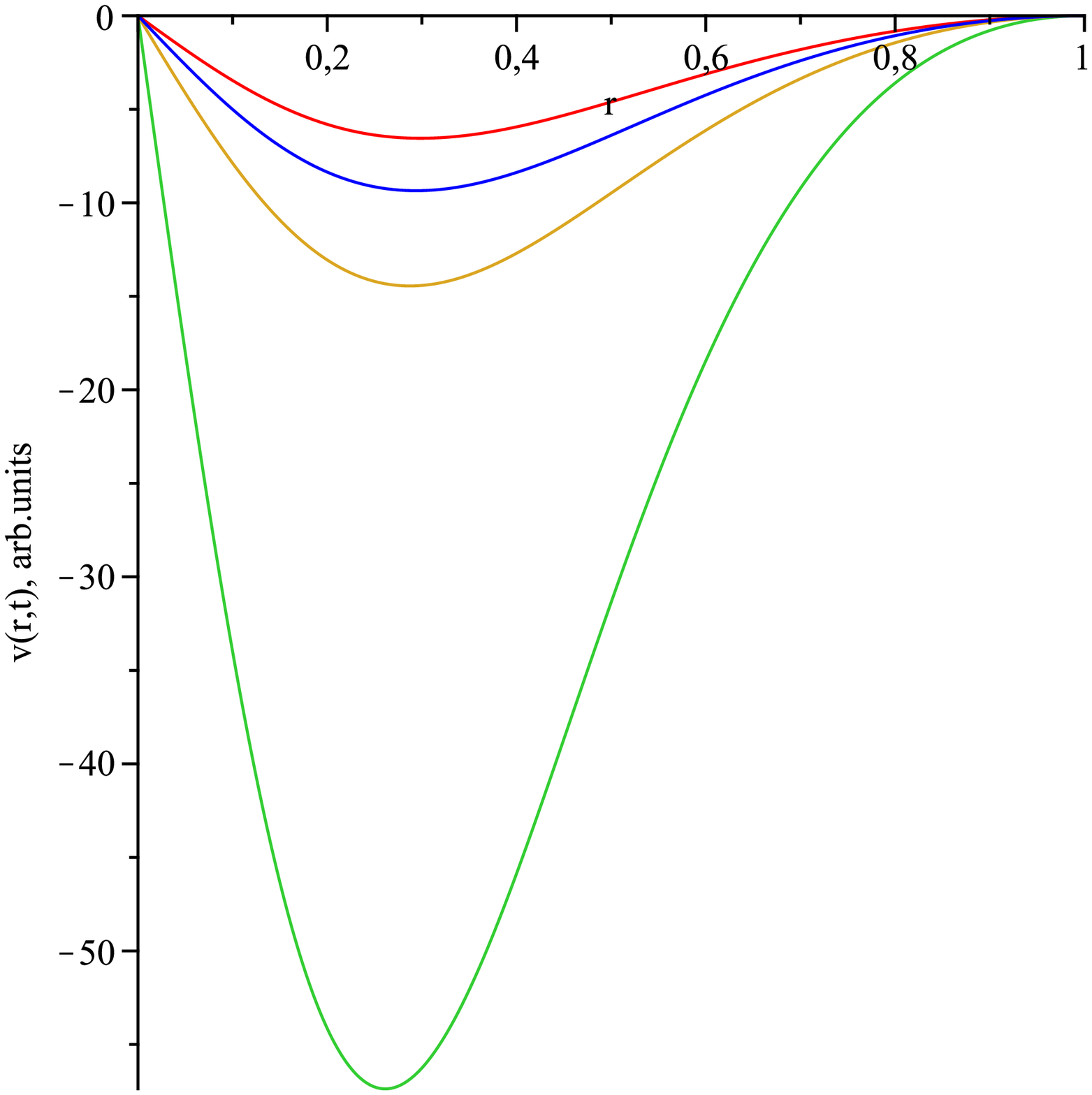}\\
\caption{Height averaged radial velocity for different modes of
evaporation: a)~\eqref{eq:Jthindisc}; b)~\eqref{eq:Jthindiscsmooth}
if $m =5$; c)~\eqref{eq:Jdavis}; d)~\eqref{eq:Jgauss}. For
additional explanations see the main text.}\label{fig:hav}
\end{figure*}

Fig.~\ref{fig:hav} demonstrates our calculations for all described
above evaporative functions. In all figures, the plots corresponds
to: initial stage (air--liquid interface is convex); air--liquid
interface is flat; air--liquid interface is concave with the same
curvature as at the initial stage; 90~\% of time, when air--liquid
interface touch the bottom of the well.

\subsection{Velocity field}\label{subsec:field}
Velocity fields calculated for sessile droplets from Laplace
equation~\cite{Tarasevich2005} and from Navier--Stokes
equations~\cite{Fischer_2002} are very similar. We hope that ideal
liquid is reasonable assumption for our object of interest.

Cylindrical coordinates $(r, \varphi, z)$ will be used, as they are
most natural for the geometry of interest. The origin is chosen in
the center of the circular well on the substrate. The coordinate $z$
is normal to the substrate, and the bottom of the circular well is
described by $z=0$, with $z$ being positive on the droplet side of
the space. The coordinates $(r,\varphi)$ are the polar radius and
the angle, respectively. Due to the axial symmetry of the problem
and our choice of the coordinates, no quantity depends on the angle
$\varphi$, in particular  $u = u(r,z)$.

The mass flux at the vertical wall of circular well is absent, hence
\begin{equation}\label{eq:boundaryrf}
    \left.\frac{\partial u}{\partial r}\right|_{r = r_f} = 0,
\end{equation}
where  $r_f$ is radius of the circular well.

Moreover, there is physically obvious relation
\begin{equation}\label{eq:boundaryr0}
    \left.\frac{\partial u}{\partial r}\right|_{r = 0} = 0.
\end{equation}

The bottom of the circular well  is impermeable, hence
\begin{equation}\label{eq:boundaryz0}
    \left.\frac{\partial u}{\partial z}\right|_{z=0} = 0.
\end{equation}
The mass flux inside the droplet near the air--liquid interface
$f(r)$ is connected with vapor flux
\begin{equation}\label{eq:boundaryzhf}
    \left.\frac{\partial u}{\partial z}\right|_\text{free surface} = f(r).
\end{equation}

Laplace equation in cylindrical coordinates for the axially
symmetric case is written as
\begin{equation}\label{eq:Laplacecyl}
    \frac{1}{r}\frac{\partial}{\partial r}\left( r \frac{\partial u}{\partial
    r}\right) + \frac{\partial^2 u}{\partial z^2} = 0.
\end{equation}

Boundary problem~\eqref{eq:boundaryrf},\eqref{eq:boundaryz0}, for
equation~\eqref{eq:Laplacecyl}
\begin{equation}\label{eq:solseries}
    u ( r, z ) = \sum\limits_{n = 1} ^ \infty a_n \cosh \left( \frac{\mu_n^{(1)}}{r_f} z
    \right) J_0 \left( \frac{\mu_n^{(1)}}{r_f} r
    \right),
\end{equation}
where  $J_m (r)$ is the Bessel function of the first kind, order
$m$, $\mu_n^{(1)}$ are the real zeros of Bessel $J_1(r)$ function:
$J_1(r) = 0$. Note that condition~\eqref{eq:boundaryr0} is satisfied
automatically.

Taking into account relation between velocity and potential $\vec{v}
= - \grad u$, we can find radial component of velocity
$$
\langle v_r(r,z) \rangle = \sum\limits_{n = 1} ^ \infty a_n \cosh
\left( \frac{\mu_n^{(1)}}{r_f} z
    \right) \frac{\mu_n^{(1)}}{r_f} J_1 \left( \frac{\mu_n^{(1)}}{r_f} r
    \right).
$$

Height averaged radial velocity, i.e. the same velocity examined
in~\cite{Rieger}, can be written as
\begin{multline*}
\langle v_r(r) \rangle= \frac{1}{h_f} \int\limits_0^{h_f} v_r(r,z)
\, dz = \\ \frac{1}{h_f} \sum\limits_{n = 1} ^ \infty a_n \sinh
\left( \frac{\mu_n^{(1)}}{r_f} h_f
    \right) J_1 \left( \frac{\mu_n^{(1)}}{r_f} r
    \right).
\end{multline*}

Coefficients $a_n$ can be obtained from~\eqref{eq:boundaryzhf}. For
simplicity we will assume that air--liquid interface is flat.
\begin{multline}\label{eq:dudz}
  \left.  \frac{\partial u}{\partial z}\right|_{z = h_f} = \\ \sum\limits_{n = 1} ^ \infty a_n \frac{\mu_n^{(1)}}{r_f} \sinh \left( \frac{\mu_n^{(1)}}{r_f}
   h_f  \right) J_0 \left( \frac{\mu_n^{(1)}}{r_f}r\right) = f(r).
\end{multline}
We introduce notation $$\int f(r) r \, dr = F(r).$$

\begin{equation}\label{eq:FBseries}
\sum\limits_{n = 1} ^ \infty c_n  J_1 \left( \frac{\mu_n^{(1)}}{r_f}
r \right) = \frac{F(r)}{r}
\end{equation}
is a Fourier--Bessel series of  $F(r)/r$,  where
\begin{equation}\label{eq:cn}
    c_n = a_n \sinh \left( \frac{\mu_n^{(1)}}{r_f} h_f \right).
\end{equation}
Hence,
\begin{multline}\label{eq:an}
    a_n  = \\ \frac{2}{r_f^2 \sinh \left( \frac{\mu_n^{(1)}}{r_f} h_f \right) J_0^2 \left(\mu_n^{(1)}
    \right)} \int\limits_0^{r_f} J_1\left( \frac{\mu_n^{(1)}}{r_f} r \right)  F(r) \, dr.
\end{multline}

We need to determine  $f(r)$ to find the unknown coefficients.

Mass conservation leads to the relation
\begin{equation}\label{eq:vzhf}
\left. \frac{\partial u}{\partial z} \right|_{z = h_f} =
-\frac{\partial L(r,t)}{\partial t} - \frac{J(r,t)}{\rho}.
\end{equation}

We computed velocity fields for all  evaporation laws described in
Sec.~\ref{subsec:hav}. The results show that current inside circular
well is horizontal excluding rather narrow region near the wall and
in the center of the well for the case of practical interest ($h/r
\sim 0.1$).

Many authors supposed that evaporation induced flow inside the
droplets is independent of its height. Our simulations confirmed
validity of this very wide used approach for thin droplets.

\begin{figure*}
\centering
 \includegraphics*[height=0.2\textheight]{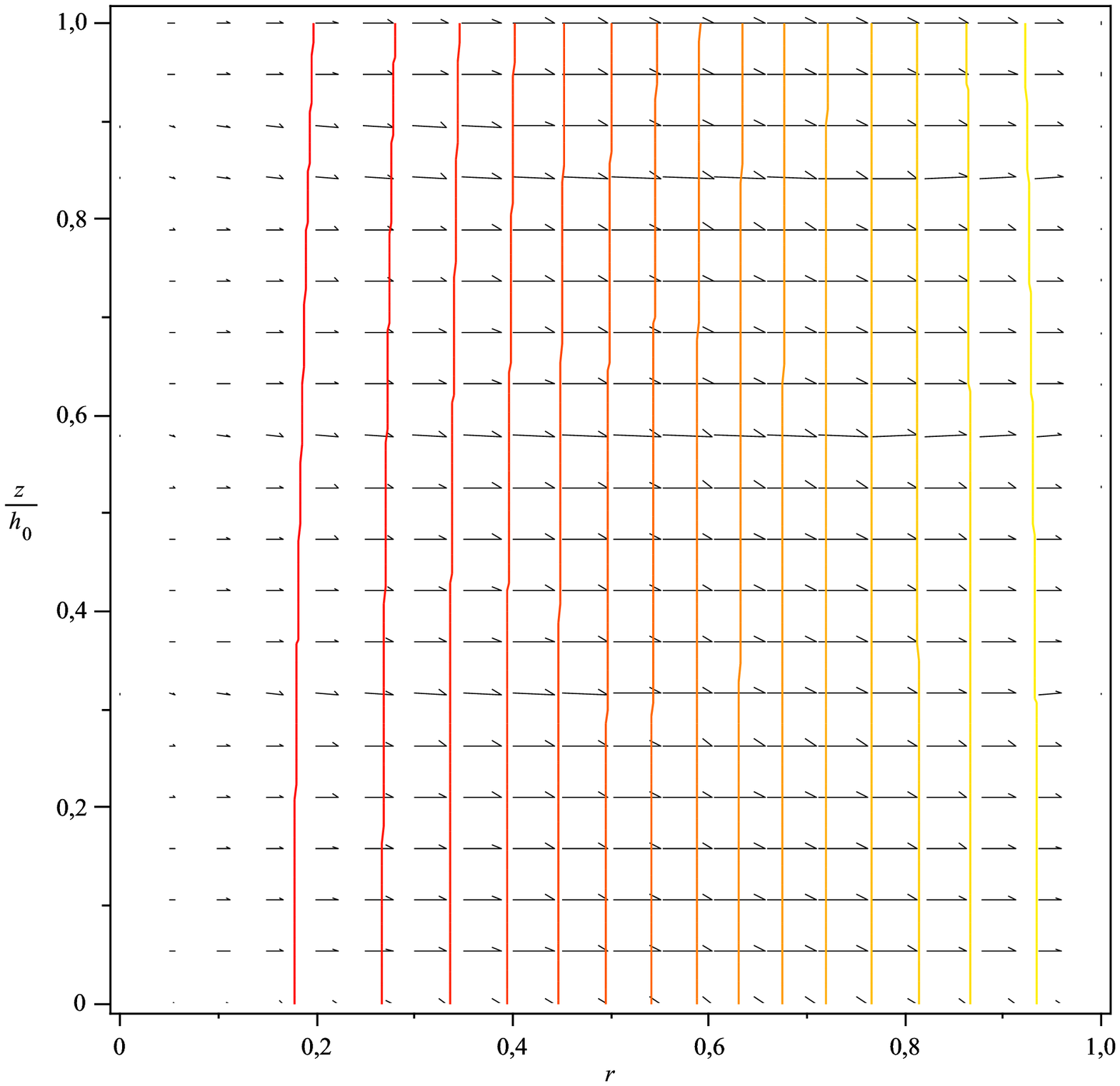} \includegraphics*[height=0.2\textheight]{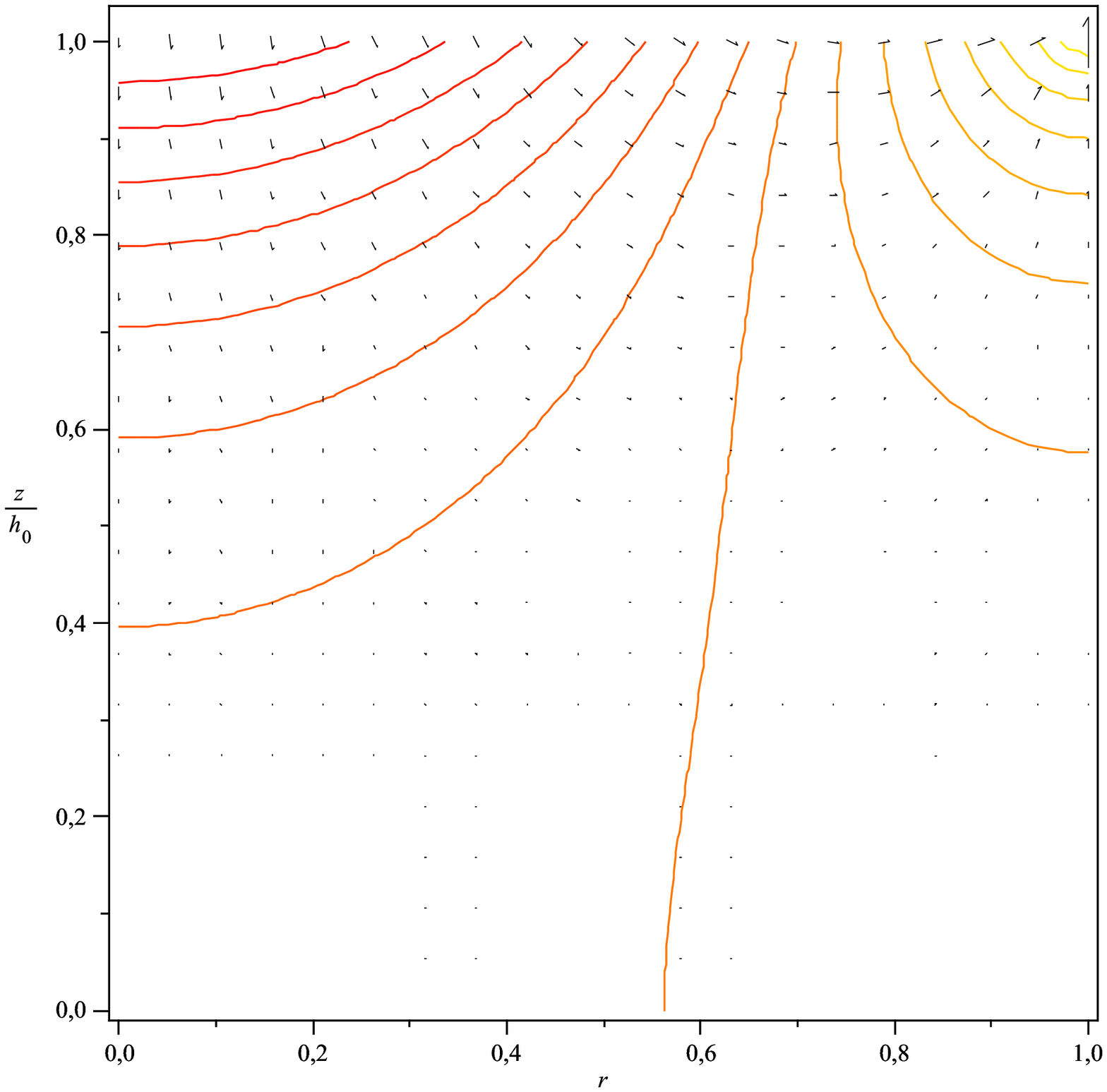} \\
 \includegraphics*[height=0.2\textheight]{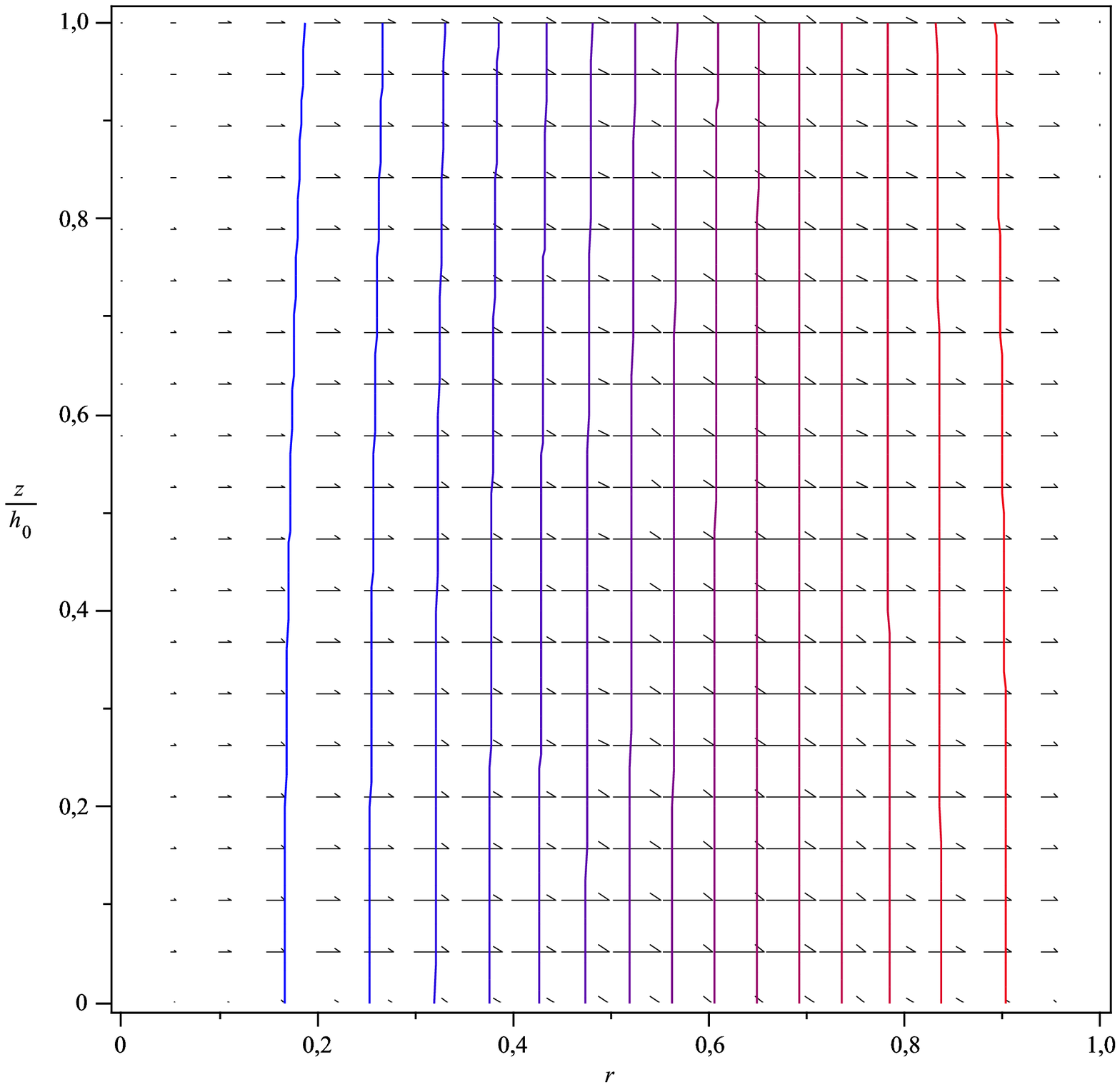}  \includegraphics*[height=0.2\textheight]{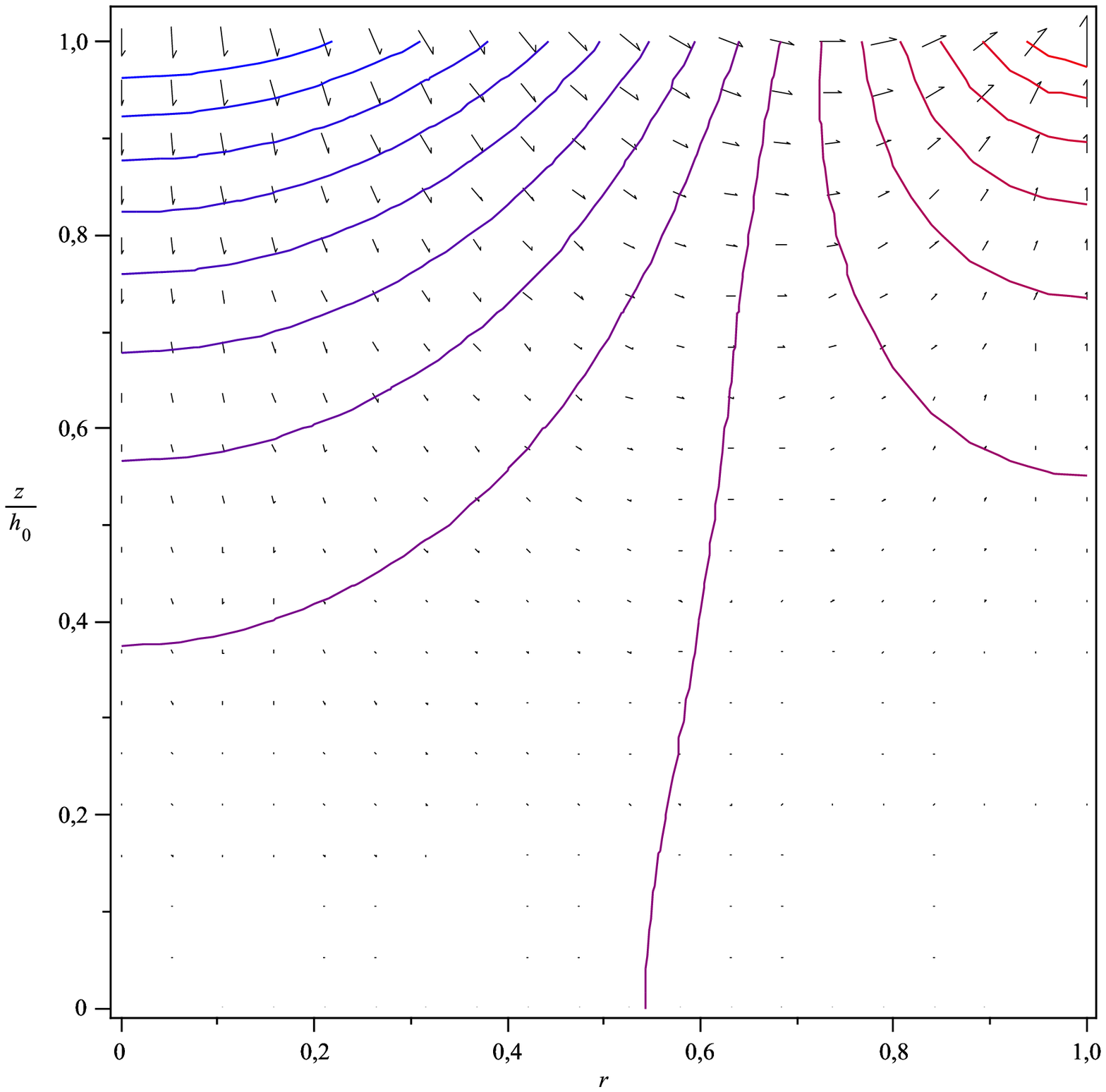}  \\
 \includegraphics*[height=0.2\textheight]{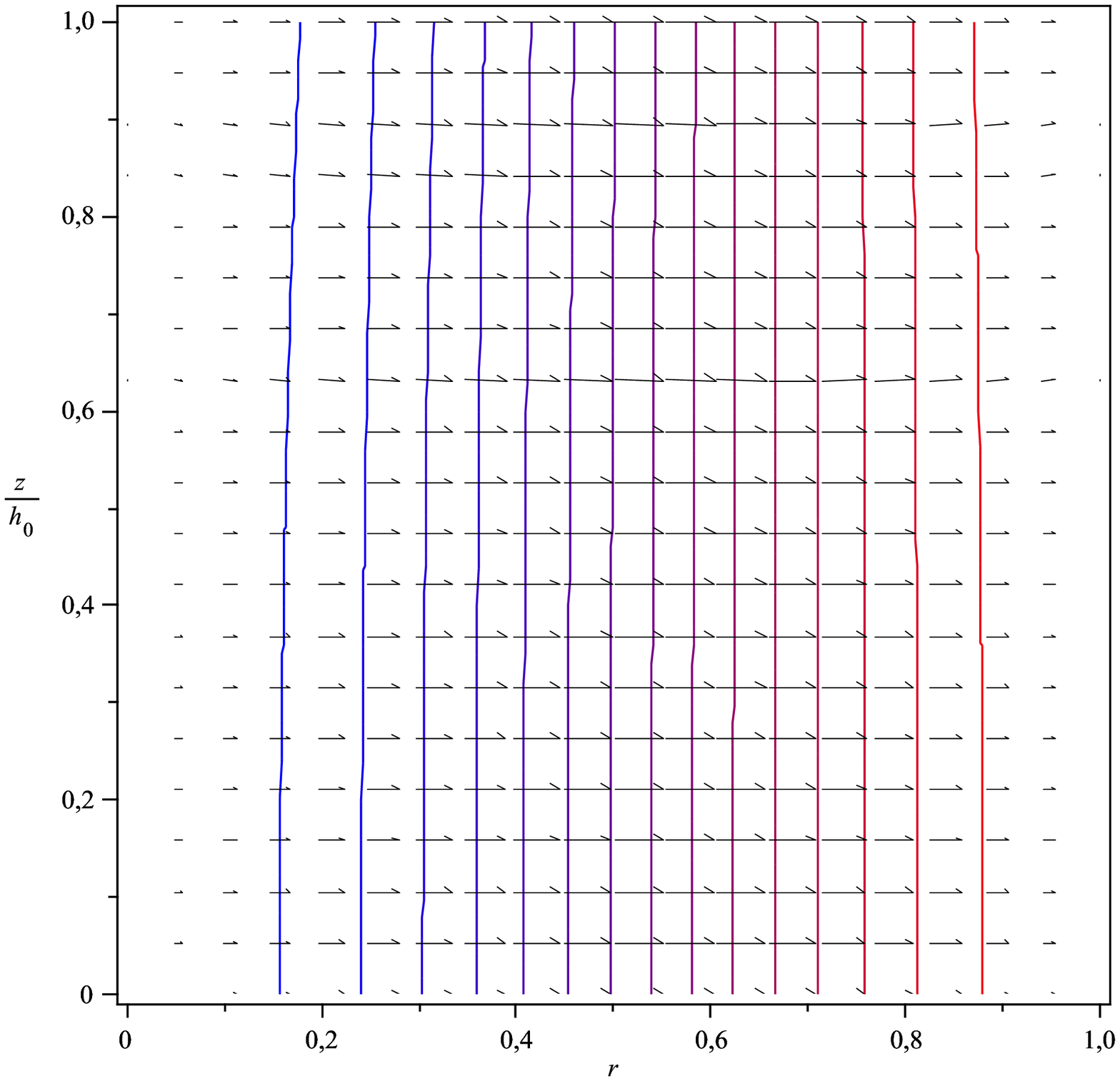}  \includegraphics*[height=0.2\textheight]{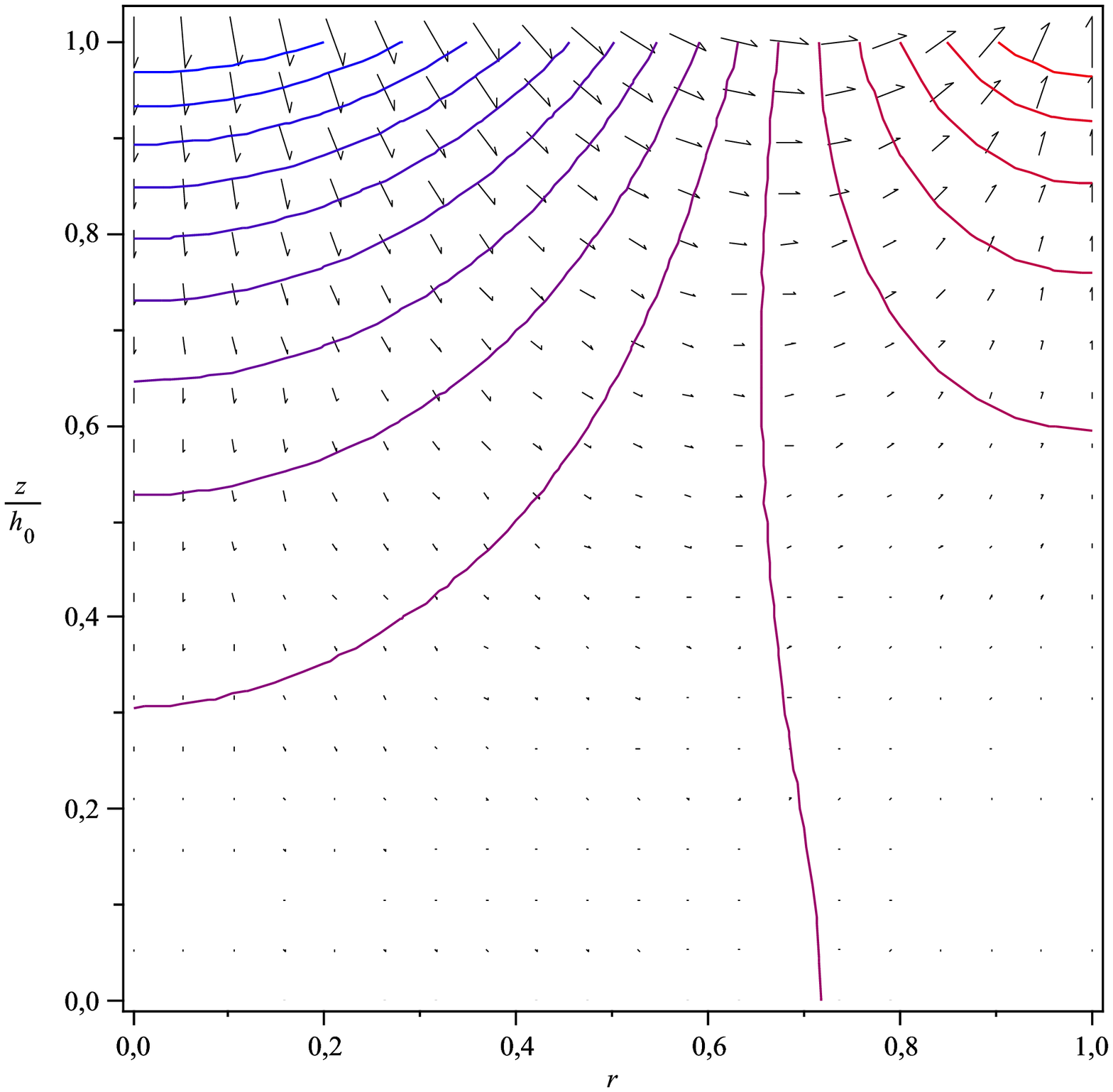}  \\
\includegraphics*[height=0.2\textheight]{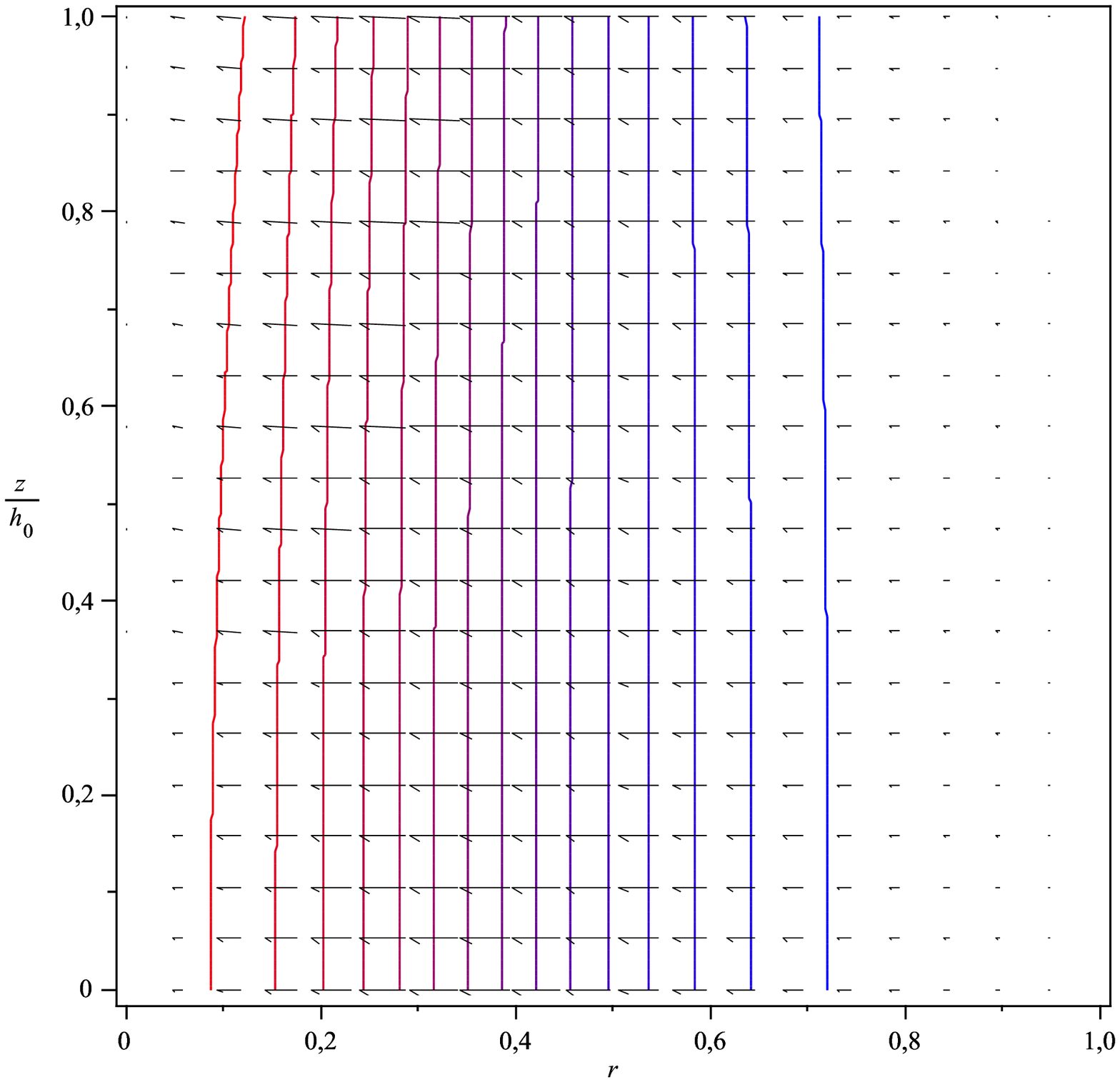}\includegraphics*[height=0.2\textheight]{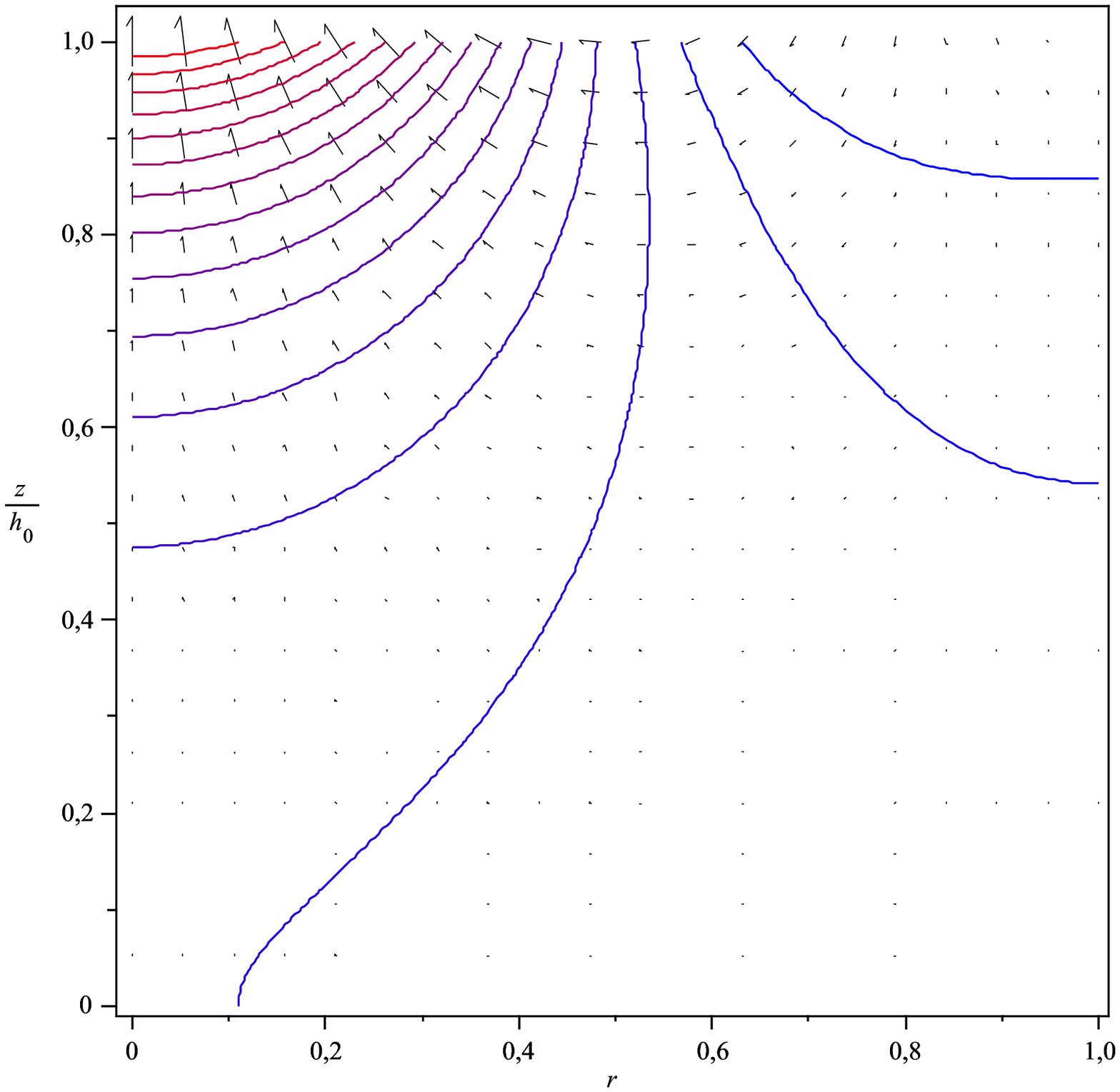}
\hfill \caption{Velocity field inside circular well for differen
evaporative modes. The air--liquid interface is supposed to be flat.
From top to bottom \eqref{eq:Jthindisc}, \eqref{eq:Jthindiscsmooth}
with $m =5$, \eqref{eq:Jdavis}, \eqref{eq:Jgauss}. Left column
corresponds to ratio $ h_f / r_f = 0.0618$, right column corresponds
to ratio $ h_f / r_f = 0.99$}\label{fig:field}
\end{figure*}

\begin{acknowledgement}
The authors are grateful to the Russian Foundation for Basic
Research for funding this work under Grant No. 06-02-16027-a.
\end{acknowledgement}


\begin{thebibliography}{99}

\bibitem{Rieger}
B.~Rieger, L.~R.~van den Doel, L.~J.~van Vliet, Physical Review E
\textbf{68}, 036312 (2003).

\bibitem{Deegan_2000}
R.~D.~Deegan, O.~Bakajin, T.~F.~Dupont, G.~Huber, S.~R.~Nagel,
T.~A.~Witten, Phys. Rev. E \textbf{62},  756 (2000)

\bibitem{Fischer_2002}
B.~J.~Fischer, Langmuir \textbf{18}, 60 (2002)

\bibitem{Mollaret_2004}
R.~Mollaret, K.~Sefiane, J.~R.~E.~Christy, D.~Veyret, Chem. Eng.
Res. Design \textbf{82(A4)}, 471 (2004)

\bibitem{Hu_2005}
H.~Hu, R.~ G.~ Larson, Langmuir \textbf{21}, 3963 (2005)

\bibitem{Popov_2005}
Y.~O.~Popov, Phys. Rev. E \textbf{71}, 036313 (2005)

\bibitem{Savina_1999}
L.~V.~Savina, \emph{Crystaloscopical structures of blood serum of
healthy and seek human}  (Sovyetskaya Kuban, Krasnodar, 1999)

\bibitem{Shabalin_2001}
V.~N.~Shabalin, S.~N.~Shatokhina, \emph{Morphology of Biological
Fluids} (Khrizostom, Moscow, 2001)

\bibitem{Cachile_2002}
M.~Cachile, O.~B\'{e}nichou, A.~M.~Cazabat, Langmuir \textbf{18},
7985 (2002)

\bibitem{Anderson_1995}
D.~M.~Anderson,  D.~M.~Davis, Physics of Fluids \textbf{7}, 248
(1995)

\bibitem{Tarasevich2005}
Y.~Y.~Tarasevich, Phys. Rev. E \textbf{71}, 027301 (2005)
\end{thebibliography}
\end{document}